\documentclass[twocolumn,secnumarabic,amssymb, nobibnotes, aps, prd]{revtex4}
\usepackage{graphicx}

\begin{document}
\title{Two permitted states of superconducting ring observed at measurements of its critical current }
\author{V.L. Gurtovoi, S.V.Dubonos, A.V.  Nikulov, and V.A. Tulin}
%\email[]{nikulov@ipmt-hpm.ac.ru}
\affiliation{Institute of Microelectronics Technology and High Purity Materials, Russian Academy of Sciences, 142432 Chernogolovka, Moscow District, RUSSIA.} %nikulov@ipmt-hpm.ac.ru
%\date{}
\begin{abstract} Magnetic dependencies of the critical current of superconducting, aluminium rings, with homogeneous section and with small narrow spot, are measured. Only permitted state at each magnetic flux value is observed in the conventional ring because of strong discreteness of its spectrum. But two states are observed because of the small narrow spot. 
 \end{abstract}

\maketitle

\narrowtext

Superconductivity is the macroscopic quantum phenomenon thanks to common momentum $p = mv + qA = \hbar \bigtriangledown \varphi$ of all superconducting pairs in a superconductor with the long-rang phase coherence $\oint_{l}dl \nabla \varphi = n2\pi $. The spectrum of a free electron in a ring with a radius $r$, like the one on the atomic orbits, should be discrete $p_{n} = n\hbar /r$ because of the Bohr' quantization $rp = n\hbar $. But quantum phenomena connected with this discreteness can be observed in real rings with $r = 1 \ \mu m$ only at very low temperatures \cite{PCnormal} because of the decrease of the energy difference between the permitted states 
$$E_{n+1} - E_{n} = \frac{mv _{n+1}^{2}}{2} - \frac{mv _{n}^{2}}{2} =  \frac{\hbar ^{2}}{2mr^{2}} (2n+1) \eqno{(1)}$$
with the radius $r$ increase. The persistent current is observed \cite{PCnormal} when the magnetic flux $\Phi $ inside the ring is not divisible $\Phi \neq n\Phi _{0}$  by the flux quantum $\Phi _{0} = 2\pi \hbar /q$ and the velocity 
$$\oint_{l}dl v_{n}  =  \frac{2\pi \hbar }{m}(n - \frac{\Phi}{\Phi_{0}})  \eqno{(2)}$$ 
can not be equal zero $v \neq 0$. In non-superconducting rings \cite{PCnormal} it is created by electrons on the Fermi level with great quantum number $n = n_{F} \approx mv_{F}r/\hbar \approx 10000$, because other electrons $n < n_{F}$ have opposite directed velocity \cite{PC1988}. Because all $ N _{s} = 2\pi rsn _{s}$ superconducting pairs in the ring with the radius $r$ and the section $s$ have the same quantum number $n$ the persistent current $I _{p} = sn _{s}qv$ is greater and their spectrum 
$$E_{n} =  N _{s} \frac{\hbar ^{2}}{2mr^{2}}(n - \frac{\Phi }{\Phi _{0}})^{2} = I _{p,A}\Phi _{0} (n - \frac{\Phi }{\Phi _{0}})^{2}  \eqno{(3)}$$
is much more discrete then the one of electrons in non-superconducting rings. $I _{p,A} = sn _{s}q (2\pi \hbar /lm)$ is the amplitude of the persistent current. The difference of the energy (3) is great $|E_{n \pm 1} - E_{n}| \gg k_{B}T$ thanks to huge number of pairs $ N _{s} $ in ring or cylinder with real sizes $r$ and $s$. Therefore the permitted state with minimum energy $\propto (n - \Phi /\Phi _{0})^{2}$ has overwhelming probability $P _{n} \propto \exp{- E_{n}/k_{B}T}$. The Little - Parks effect \cite{LP1962} corroborates this overwhelming probability even in the fluctuation region above superconducting transition $T > T_{c}$. According to the universally recognized explanation \cite{LP1962,Tinkham} the oscillations of the resistance $R(\Phi /\Phi _{0})$ of superconducting cylinder or ring at $T \geq  T_{c}$ is a consequence of the change of the quantum number $n$ corresponding to lowest energy $\propto (n - \Phi /\Phi _{0})^{2}$ at $\Phi = (n+0.5)\Phi _{0}$. 

\begin{figure}[b]
\includegraphics{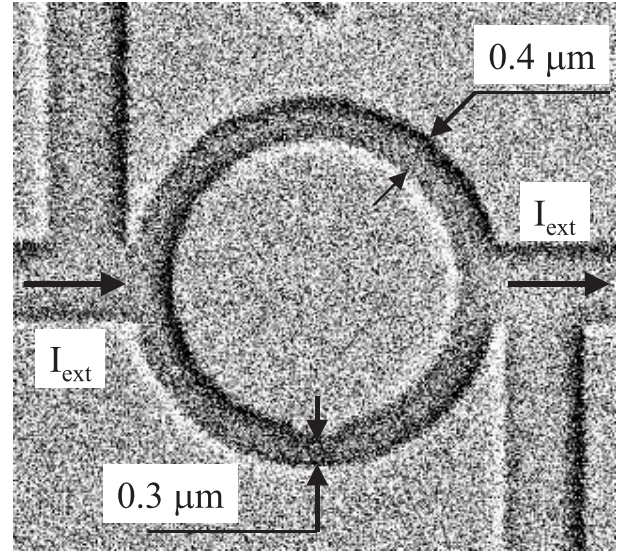}
\caption{\label{fig:epsart} The ring with small narrow spot measurement of the magnetic dependencies of the critical current of which has demonstrate the observation of the two permitted states $n $ and $n \pm 1$. The ring radius $r \approx  2 \ \mu m$. }
\end{figure}

The strong discreteness $E_{n+1} - E_{n} \gg k_{B}T$ of the permitted state spectrum of superconducting ring is corroborated also with help of measurements of the critical current $I_{c}$. Its value $I_{c}(\Phi /\Phi _{0})$ corresponds to the measuring current $|I _{ext}|$, Fig.1, at which the superconducting pair velocity reaches the depairing velocity $v _{c}$ in one of the ring halves and superconducting state is broken by jump \cite{Eprint2006}. The pair velocity in the ring halves $v _{up}$ and $v _{down}$ are determined with the measuring current $I _{ext} = I _{up} + I _{down} = q(s_{up}n_{s,up}v_{up} + s_{ down }n_{s,down}v_{down})$ and the quantization (2) according to which $ l_{up}v_{up} - l_{ down}v_{down} =  (2\pi \hbar /m)(n - \Phi / \Phi_{0}) $. In a symmetric ring, with the same parameters of the ring halves $s_{up} = s_{down}$, $n_{s,up} = n_{s,down}$, $l_{up} = l_{down} = l/2$ the velocities should be equal $ v_{up} = I _{ext}/2sn _{s}q + (2\pi \hbar /lm)(n - \Phi / \Phi_{0}) = (I _{ext} + 2I _{p})/2sn _{s}q $, $v_{down} = I _{ext}/2sn _{s}q - (2\pi \hbar /lm)(n - \Phi / \Phi_{0}) = (I _{ext} - 2I _{p})/2sn _{s}q $ and the critical current determined with the condition $|v _{up}| = v _{c}$ or $|v _{down}| = v _{c}$ 
$$I _{c} = I _{c,0} - 2|I _{p}| = I _{c,0} - 2I _{p,A}2| n - \frac{\Phi}{\Phi_{0}}|  \eqno{(4)}$$
should depend on the persistent current $ I _{p} = I _{p,A} 2(n - \Phi / \Phi_{0})$ value. $ I _{c,0} = sn _{s}q v _{c}$ is the critical current at $ I _{p} = 0$. As positive, the direction from right to left for $I _{ext}$, $v_{up}$, $v_{down}$ and clockwise for the persistent current $I _{p}$ are taken. 

\begin{figure}
\includegraphics{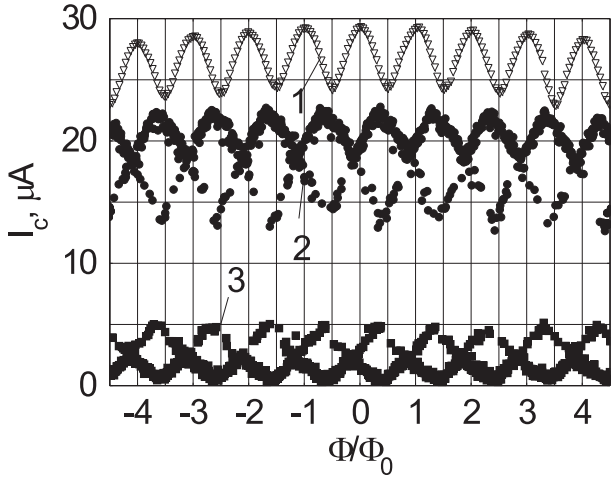}
\caption{\label{fig:epsart} The magnetic dependencies of the critical current measured on the conventional ring with the radius $r \approx  2 \ \mu m$ (1) and on the like ring but with small narrow spot, see Fig.1, (2). The dependence of the absolute value of the persistent current $|I _{p}| = (I _{c,0} - I _{c})/2$ corresponding, according (4), to the $I _{c}(\Phi / \Phi_{0})$ dependence measured on the ring with small narrow spot is shown also (3). }
\end{figure}

The magnetic dependencies of the critical current measured on the conventional, symmetric ring are described very well by the relation (4) in which the quantum number changes from $n$ to $n+1$ at $\Phi = (n+0.5)\Phi _{0}$, Fig.2. The amplitude $6 \ \mu A$ of the oscillations of the critical current of the conventional ring shown on Fig.2 corresponds, according to (4), the amplitude of the persistent current $ I _{p,A} \approx  3 \ \mu A$.  The energy difference $ E_{n \pm 1} - E_{n} \approx  I _{p,A} \Phi _{0} \approx  6 \ 10^{-21} \ J$, see (3), between the lowest state $n$ and the adjacent state $n \pm 1$, at this $ I _{p,A} $ corresponds to the temperature $ (E_{n \pm 1}- E_{n}) / k _{B} \approx 450 \ K$ exceeding strongly the temperature of measurements $T _{m} \approx  1.45 \ K \approx  0.97T _{c} $, Fig.2. Therefore the critical current (4) corresponding to only permitted state $n$ is observed at $n - 0.5 < \Phi / \Phi_{0} < n + 0.5$ in accordance with the enormous ratio of the probabilities $P _{n}/ P _{n \pm 1} = \exp{(E_{n \pm 1}- E_{n})/k_{B}T} \approx  e^{310} \approx  10^{134}$ of the transition of the ring in superconducting state with the quantum number $n$ and $n \pm 1$. But the measurements have revealed that even small narrow spot in a ring section, Fig.1, can increase strongly the probability of the $n \pm 1$ state. According to the result of the $I _{c}$ measurements shown on Fig.2 the probability ratio $P _{n}/ P _{n \pm 1}$ does not exceed 3. This result should imply the small energy difference $E_{n \pm 1}- E_{n} \approx  \ln {3} \ k_{B}T _{m} \approx  k_{B} \ 1.6 \ K$.  

In additional to the increase of the $n \pm 1$ state probability the narrow spot in one of the ring halves results to a shift of the $I _{c}(\Phi / \Phi_{0})$ dependencies on $0.33 \Phi_{0}$, Fig.1. This paradoxical effect was investigated in detail at measurements of the critical current of rings with different sections of ring halves \cite{Eprint2006}. The shift is observed because of an asymmetry, i.e. a distinction between section of ring halves \cite{Eprint2006}. The two states $n$ and $n \pm 1$ can be observed without the shift of the $I _{c}(\Phi / \Phi_{0})$ dependencies at measurement of the critical current of a ring with two symmetric narrow spots in ring halves \cite{2states2002}. The shift results to the obvious contradiction of the $I _{c}(\Phi / \Phi_{0})$ dependencies measured on the asymmetric ring with the prediction of the universally recognised quantum formalism (2) and the Little - Parks resistance oscillations measured on the same asymmetric ring \cite{Eprint2006}. According to the quantization requirement (2) extreme values must be observed at $\Phi = n\Phi_{0}$ and $\Phi = (n+0.5)\Phi_{0}$. Maximums and minimums of the critical current oscillations $I _{c}(\Phi / \Phi_{0})$ of symmetric rings, Fig.2, as well as the resistance oscillations $R(\Phi / \Phi_{0})$ of asymmetric rings \cite{Eprint2006} are observed in accordance with this requirement. But the observation of the extreme values of the $I _{c}(\Phi / \Phi_{0})$ dependencies of asymmetric rings at $\Phi = (n+0.35)\Phi_{0}$ and $\Phi = (n+0.85)\Phi_{0}$, Fig.2, contradicts inexplicably to the quantization condition (2). Therefore the experimental results shown on Fig.2 can not be explain completely.    

We can assume only that the observation of the two states in a symmetric ring with two narrow spots in ring halves may be connected with non-monotonic dependence $j_{s} = qn _{s}v = qn _{s0}(1 - v^{2}/3v _{c}^{2})v$ of the superconducting current density $j_{s}$ on pair velocity $v$ \cite{Tinkham}. Superconducting state with $3 v _{c} > v _{sh} > v _{c}$ and the reduced pair density $n _{s,sh} = n _{s0}(1 - v^{2}/3v _{c}^{2}) \ll n _{s0}$ can be stable in a short ring segment $l _{sh} < \xi (T)$ thanks to the proximity effect when $v \ll v _{c}$ and $n _{s} \approx  n _{s0}$ in the adjacent ring segment. The short ring segment can give a considerable contribution $l _{sh} v _{sh} \approx  \pm2 \pi \hbar /m $ to the velocity circulation (2) and a small one $s _{sh}l _{sh}n _{s,sh} mv _{sh}^{2}/2 = l _{sh} mv _{sh}I _{p}/2$ in the energy (3). Therefore the inhomogeneous state with $ n _{s,sh} \ll n _{s0}$ in the short ring segment $l _{sh} $ and  $n _{s} \approx  n _{s0}$ in the rest of the ring can have the quantum number $n \pm 1$ and the energy differs to a little degree from the energy of the homogeneous state with the quantum number $n$. 

\section*{Acknowledgement}
This work has been supported by a grant "Quantum bit on base of micro- and nano-structures with metal conductivity" of the Program "Technology Basis of New Computing Methods" of ITCS department of RAS, a grant of the Program "Low-Dimensional Quantum Structures" of the Presidium of Russian Academy of Sciences and a grant 04-02-17068 of the Russian Foundation of Basic Research.

\end{document}